\documentclass[a4paper]{revtex4}
\usepackage{graphicx}
\usepackage{fancyhdr}
\usepackage{amsmath}
\usepackage{color}

\begin{document}

\title{The One Higgs and its Connections}

%

\author{Ernest Ma}
\affiliation{Physics and Astronomy Department, Uniersity of California, 
Riverside, CA 92521, USA}

\begin{abstract}
I propose the notion that there is only one Higgs doublet which is 
responsible for electroweak symmetry breaking and all other possible 
scalars in any extension of the standard model are prevented from mixing  
with it because of a symmetry which stabilizes dark matter.  This leads 
naturally to radiative (scotogenic) neutrino mass as well as radiative 
quark and lepton masses if flavor symmetry is also considered.
\end{abstract}

\maketitle

\thispagestyle{fancy}


\section{One Higgs Does All?}
With the discovery of the 125 GeV particle~\cite{atlas12,cms12} at CERN in 
2012, and the absence of any other evidence of new physics, together with 
subsequent data confirming that it is indeed consistent with being the one 
predicted Higgs boson of the Standard Model (SM), it has become essential 
to ask the question:  Is that it?

After all, the SM needs just the one Higgs.  It is capable of doing it all, 
i.e. gives masses to the $W^\pm$ and $Z$ bosons, as well as all the quarks 
and leptons.  This means that the SM is potentially complete, and there is 
nothing else fundamental to discover, excepting of course the origins of 
neutrino mass and dark matter (DM).

On the other hand, the discovered 125 GeV particle may not be exactly the 
one Higgs predicted by the SM.  It may still hold some surprises as its 
properties are being scrutinized experimentally.  On the phenomenological 
side, it opens up the new possibility to use it as a probe of new physics, 
which is the topic of this workshop.  On the theoretical side, we may want 
to understand how this one Higgs occurs in a natural extension of the SM.

\section{More Higgs or One Higgs Does More?}
Numerous studies have been made on the extensions of the scalar sector of the 
SM: PACS category 12.60.Fr.  In particular, two Higgs doublet models abound: 
Type I, II, X, Y, etc.  In all such cases, the linear combination with 
$\langle \phi^0 \rangle = 174$ GeV, i.e. that of the SM, is in general not 
automatically a mass eigenstate.

Another often studied scenario is that of flavor symmetry.  To realize such 
a symmetry in a renormalizable theory, the Higgs Yukawa couplings 
$f_{ijk} \bar{Q}_{iL} d_{jR} \Phi_k$, etc. require that there be more than one 
Higgs doublet.
What if there is only One Higgs?, i.e. a scalar doublet with 
$\langle \phi^0 \rangle = 174$ GeV, which is prevented from mixing with any 
other scalar by an approximate or exactly conserved symmetry.

The former may be realized in a model of of flavor~\cite{mm13} for example. 
Here under the symmetry $S_3 \times Z_2$, there are three Higgs doublets: 
$(\Phi_1,\Phi_2) \sim (2,+)$ and $\Phi_3 \sim (1,-)$.  The $S_3 \times Z_2$ 
symmetry breaks softly to $Z_2 \times Z_2$, both of which are then also 
softly broken.  The $(-,+)$ Higgs mediates $B - \bar{B}$ mixing with a mass 
$\sim 1$ TeV.  The $(+,-)$ Higgs mediates $K - \bar{K}$ mixing with a mass 
$\sim 10$ TeV.  The $(+,+)$ Higgs is almost exactly that of the SM.  There 
is a unique prediction of this model, i.e. the branching fraction of 
$B_s \to \tau^+ \mu^-$ may be as high as $10^{-7}$, whereas that of $B \to 
\mu^+ \tau^-$ is much more suppressed.  Note that $B_s \to \mu^+ \mu^-$ has 
been seen~\cite{lhcb14} with a branching fraction of $2.8 ~(+0.7/-0.6) 
\times 10^{-9}$.

The latter was the inspiration of the scotogenic model~\cite{m06} of radiative 
neutrino mass (from the Greek {\it scotos} meaning darkness, which identifies 
this symmetry as $Z_2$ for dark matter.  A second scalar doublet 
$(\eta^+,\eta^0)$ and three neutral fermion singlets $N_{1,2,3}$ are added to 
the SM which are odd under $Z_2$, whereas all other (i.e. SM) particles are 
even.  The complex scalar $\eta^0 = (\eta_R + i \eta_I)/\sqrt{2}$ is split 
so that $m_R \neq m_I$.  Let $m_R < m_I$, then $\eta_R$ is a dark-matter 
candidate.  Alternatively, it may be the lightest $N$.  Using 
$f(x) = -\ln x/(1-x)$, the Majorana neutrino mass matrix is given by
\begin{equation}
({\cal M}_\nu)_{\alpha \beta} = \sum_i {h_{\alpha i} h_{\beta i} M_i \over 16 \pi^2} 
[f(M_i^2/m_R^2) - f(M_i^2/m_I^2)].
\end{equation}
Dark matter is either WIMP (Weakly Interacting Massive Particle) in a 
freeze-out scenario, or FIMP (Feebly Interacting Massive Particle) in a 
freeze-in scenario.
\begin{figure}[htb]
\vspace*{-3cm}
\hspace*{-3cm}
\includegraphics[scale=1.0]{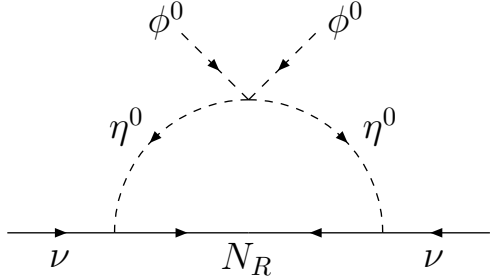}
\vspace*{-21.5cm}
\caption{One-loop generation of neutrino mass with $Z_2$ symmetry.}
\end{figure}

\section{The Dark Matter and Flavor Connection}
The Higgs connection to $W,Z$ bosons is fundamental to the SM and should not 
be changed.  {\it Question}:  What about its other connections?  It has 
already been conjectured that it connects to neutrinos only through DM. 
Why not to (some) quarks and leptons?  This is the subject of my recent 
paper~\cite{m14}: Instead of the Higgs boson coupling directly to fermions, a 
dark $U(1)_D$ symmetry as well as a flavor symmetry (such as $A_4$) are 
imposed to forbid certain Yukawa couplings.  The flavor symmetry is then 
softly broken and the fermion gets a radiative scotogenic mass 
in one loop.  The $U(1)_D$ symmetry may also be broken into a residual $Z_2$ 
symmetry.  Examples are shown below.
\begin{figure}[htb]
\vspace*{-3cm}
\hspace*{-3cm}
\includegraphics[scale=1.0]{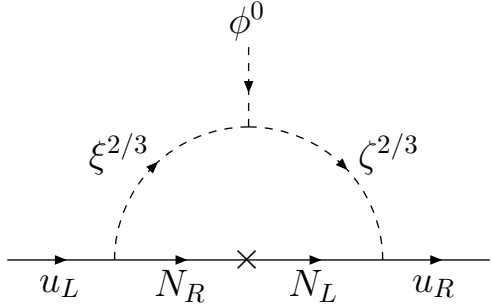}
\vspace*{-21.5cm}
\caption{One-loop generation of $u$ quark mass with $U(1)_D$ symmetry.}
\end{figure}
\begin{figure}[htb]
\vspace*{-3cm}
\hspace*{-3cm}
\includegraphics[scale=1.0]{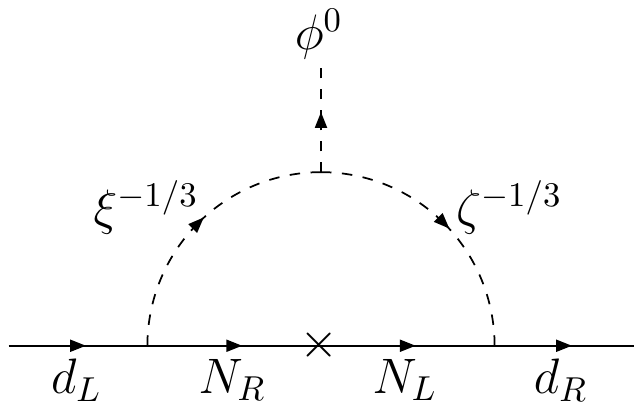}
\vspace*{-21.5cm}
\caption{One-loop generation of $d$ quark mass with $U(1)_D$ symmetry.}
\end{figure}
\begin{figure}[htb]
\vspace*{-3cm}
\hspace*{-3cm}
\includegraphics[scale=1.0]{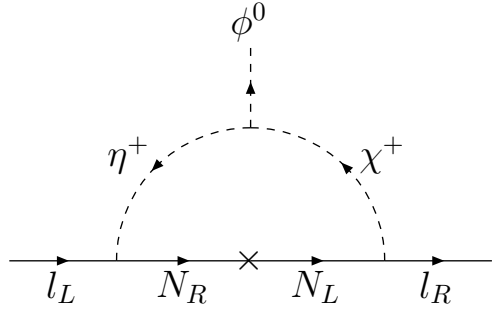}
\vspace*{-21.5cm}
\caption{One-loop generation of charged-lepton mass with $U(1)_D$ symmetry.}
\end{figure}
In these diagrams, there appear the analogs of scalar leptons and quarks as 
in supersymmetry, but the model is not supersymmetric.  Note that 
$(\xi^{2/3},\xi^{-1/3})$ is a scalar electroweak doublet, whereas $\zeta^{2/3}$ 
and $\zeta^{-1/3}$ are singlets.  There is also only one copy of these scalars, 
instead of three copies in supersymmetry because there are three families 
of quarks.  The flavor information is carried by the $N_{1,2,3}$ singlet 
fermions.  The $3 \times 3$ ${\cal M}_N$ mass matrix softly breaks $A_4$ 
and its structure is transmitted to the quark and lepton mass matrices 
through their different couplings to the dark sector.  In this way, a 
unified understanding of quark and lepton mixing is obtained by tracing 
its origin to the structure of dark matter.

\section{Anomalous Higgs Yukawa Couplings}
An immediate consequence of this scenario is a possible observable 
deviation~\cite{fm14} of the Higgs Yukawa coupling of $h \bar{\psi} \psi$ 
which is well-known to be given by $m_\psi/v$ in the SM where $v = 246$ GeV. 
In the radiative mechanism for leptons, the doublet $(\eta^+,\eta^0)$ and 
singlet $\chi^+$ mix through the term $\mu(\eta^+ \phi^0 - \eta^0 \phi^+) 
\chi^-$, where $\langle \phi^0 \rangle = v/\sqrt{2}$.  Let the mass 
eigenstates be $\zeta_1 = \eta \cos \theta + \chi \sin \theta$, and 
$\zeta_2 = \chi \cos \theta - \eta \sin \theta$ with masses $m_1$ and 
$m_2$, then $\mu v/\sqrt{2} = \sin \theta \cos \theta (m_1^2 - m_2^2)$. 
Let $x_{1,2} = m_{1,2}^2/m_N^2$, the one-loop mass is
\begin{equation}
m_l = {f_\eta f_\chi \sin \theta \cos \theta m_N \over 16 \pi^2} 
\left( {x_1 \ln x_1 \over x_1 - 1} - {x_2 \ln x_2 \over x_2 - 1} \right).
\end{equation}
\begin{figure}[htb]
\vspace*{0.5cm}
\includegraphics[scale=1.32]{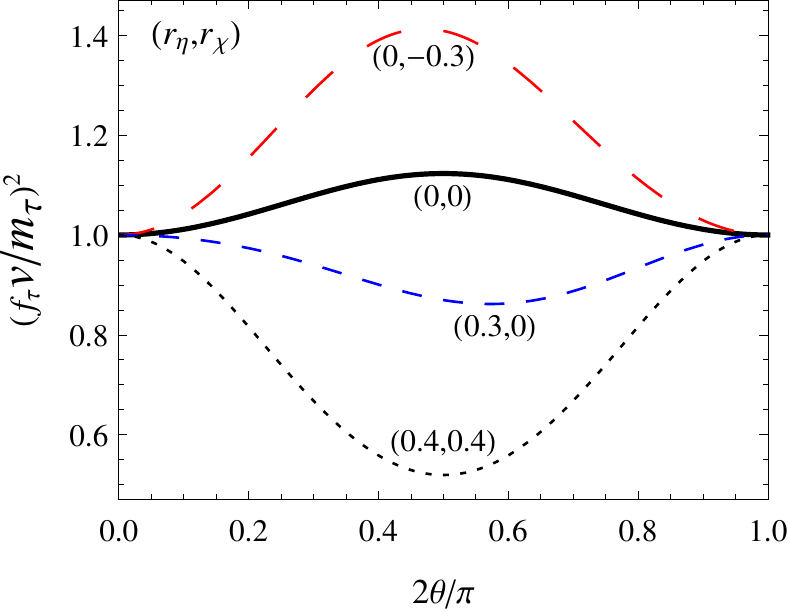}
\caption{The ratio $(f_\tau v/m_\tau)^2$ plotted against $\theta$ for 
$x_1 = 3$ and $x_2 = 1$ with various $(r_\eta,r_\chi)$.}
\end{figure}
The Yukawa coupling of $h$ to $\bar{l} l$ is now not exactly equal to 
$m_l/v$.  It has three contributions, through $\eta^+ \eta^-$, $\chi^+ \chi^-$, 
and $\eta^\pm \chi^\mp$.  Let $r_{\eta,\chi} = \lambda_{\eta,\chi} (m_N/\mu)^2$, 
then
\begin{equation}
{f_l v \over m_l} = 1 + {1 \over 2} (\sin 2 \theta)^2 (a_+ F_+ + a_- F_-),
\end{equation}
where
\begin{eqnarray}
&& a_+ = 1 + (x_1-x_2) \cos 2 \theta (r_\eta - r_\chi), ~~~~ 
a_- = (x_1 - x_2)(r_\eta + r_\chi), \\ 
&& F_+ = [F(x_1,x_1) + F(x_2,x_2)]/2F(x_1,x_2) - 1, ~~~~  
F_- = [F(x_1,x_1) - F(x_2,x_2)]/2F(x_1,x_2),  
\end{eqnarray}
with
\begin{equation}
F(x_1,x_2) = {1 \over x_1-x_2} \left( {x_1 \ln x_1 \over x_1 - 1} - 
{x_2 \ln x_2 \over x_2 - 1} \right).
\end{equation}
Note that the deviation from 1 is not suppressed by the usual loop factor of 
$16 \pi^2$ as in other radiative-correction calculations.
\begin{figure}[htb]
\vspace*{0.5cm}
\includegraphics[scale=1.32]{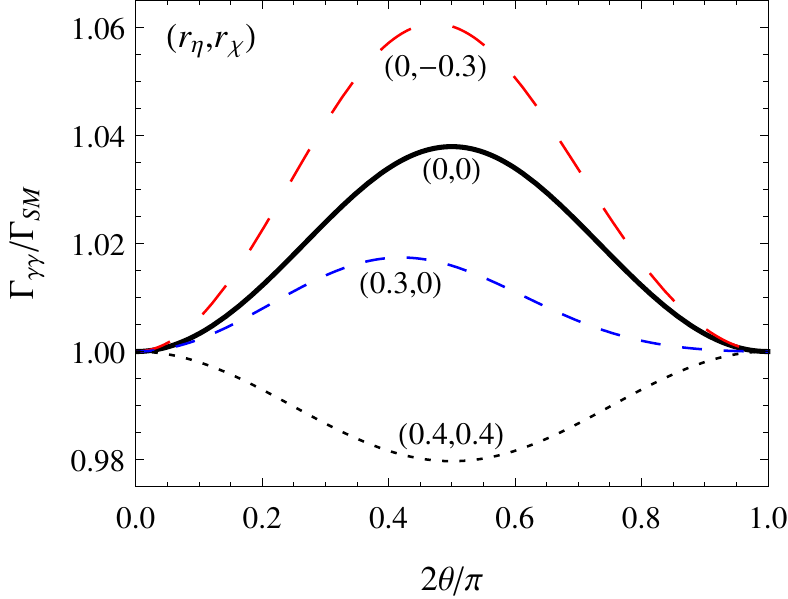}
\caption{The ratio $\Gamma_{\gamma \gamma}/\Gamma_{SM}$ plotted against 
$\theta$ for $x_1=3$ and $x_2=1$ with various $(r_\eta,r_\chi)$ and 
$\mu/m_N =1$.}
\end{figure}
\begin{figure}[htb]
\vspace*{-0.5cm}
\includegraphics[scale=1.28]{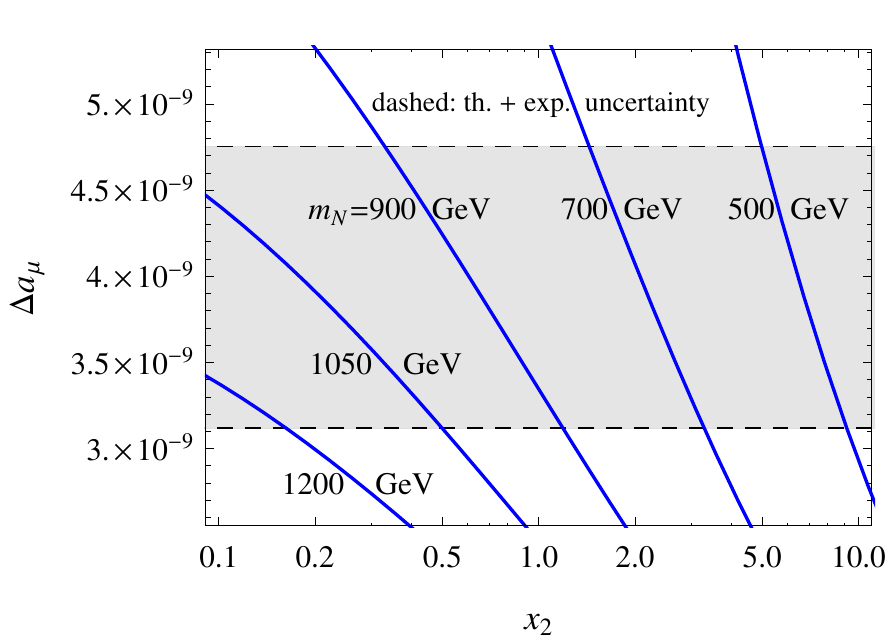}
\caption{$\Delta a_\mu$ plotted against $x_2$ with $x_1 = x_2 +2$ 
for various $m_N$.}
\end{figure}
Take for example $x_1=3$ and $x_2=1$, then $F(3,1) = 0.324$.  For $m_\tau$, this 
yields $f_\eta f_\chi/4 \pi = 0.4(m_N/\mu)$ and $\sin 2 \theta = \mu v/\sqrt{2} 
m_N^2$.  Hence $m_N > 174$ GeV for $m_N/\mu < 1$.  The effect on 
$(f_\tau v/m_\tau)^2$ is plotted for various values of $(r_\eta,r_\chi)$ in 
Fig.~5.
The charged scalars $\zeta_{1,2}$ also contribute to $h \to \gamma \gamma$. 
Assuming again $x_1=3$ and $x_2=1$, and also $\mu/m_N=1$, $\Gamma_{\gamma \gamma}/
\Gamma_{SM}$ is plotted against $\theta$ in Fig.~6.
If we apply the same procedure to the muon, then
\begin{equation}
\Delta a_\mu = {(g-2)_\mu \over 2} = {m_\mu^2 \over m_N^2} \left[ 
{G(x_1) - G_(x_2) \over H(x_1) - H(x_2)} \right],
\end{equation}
where
\begin{equation}
G(x) = {2x \ln x \over (x-1)^3} - {x+1 \over (x-1)^2}, ~~~ H(x) = 
{x \ln x \over x-1}.
\end{equation}
This is plotted against $x_2$ with $x_1 = x_2 + 2$ for various $m_N$ in Fig.~7. 
Note that the usual factor of $16 \pi^2$ is missing in the denominator, thus 
allowing for a larger value of $m_N$ to explain the experimental deviation 
of $\Delta a_\mu$ from the SM prediction.

%



\section{Collider Signatures}
The dark-matter singlet neutral fermions $N_{1,2,3}$ carry flavor and their 
mixing pattern gets transmitted to the quarks, leptons, and neutrinos.  
Since flavor is organized through them, the production of scalar quarks, 
then $\tilde{q} \to q_{1,2} N_{1,2}$ with $N_2 \to \eta/\chi + \mu^\pm$ and 
$\eta/\chi \to N_1 e^\mp$ will result in
$$2~{\rm jets} + \mu^\pm + e^\mp + {\rm missing~energy}$$ 
at the Large Hadron Collider (LHC).  In contrast, in the Minimal Supersymmetric 
Standard Model, the neutral gauginos do not carry flavor, so the decay of 
squarks to dark matter will mostly result in 2 jets + $\mu^+ \mu^-$ (or 
$e^+ e^-$) + missing energy instead.  Consider the expected 13 TeV run at 
the LHC.  This signature is observable~\cite{mn15} with $S/N > 5$ where the 
SM background is mainly $t \bar{t}$ production.  
\begin{figure}[htb]
\includegraphics[scale=1]{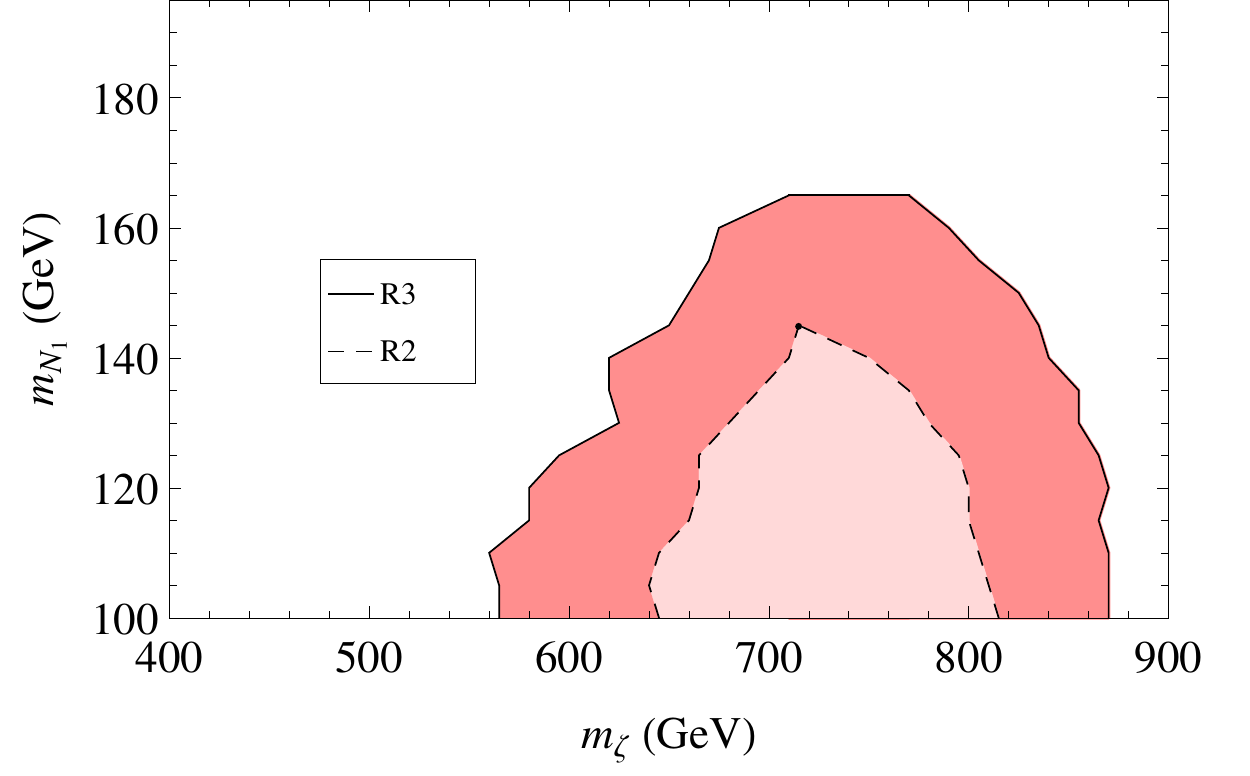}
\caption{Masses for $N_1$ and $\zeta$ that could produce a 
signal-to-background ratio, when compared to $ t \bar{t}$ decays, larger 
than 5 in the opposite-sign opposite-flavor dilepton + 2 jets + missing 
energy signature, under the R2 and R3 cuts.}
\end{figure}
\begin{figure}[htb]
\includegraphics[scale=1]{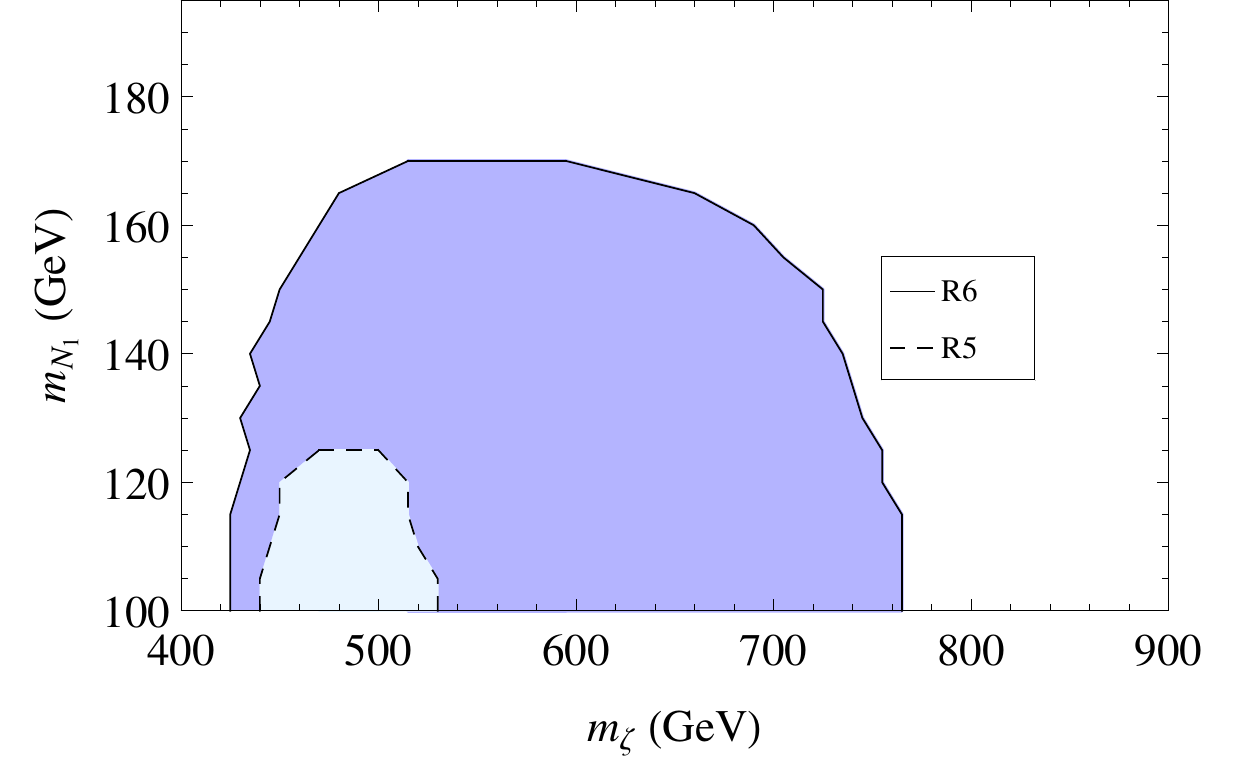}
\caption{Masses for $N_1$ and $\zeta$ that could produce a 
signal-to-background ratio, when compared to $ t \bar{t}$ decays, larger 
than 5 in the opposite-sign opposite-flavor dilepton + 2 jets + missing 
energy signature, under the R5 and R6 cuts.}
\end{figure}
Applying the cuts: 
$|\eta_j| < 3$, $|\eta_e| < 2.4$, and $|\eta_\mu| < 2.5$, we consider four 
cut regions with $m_{N_2} = 400$ GeV and $m_\chi = 200$ GeV:
\begin{eqnarray}
R2&:& T_T^m > 200~{\rm GeV}, ~~H_T > 600~{\rm GeV}, ~~p_T^j > 30~{\rm GeV}, 
~~ p_T^l > 20~{\rm GeV}; \\ 
R3&:& T_T^m > 275~{\rm GeV}, ~~H_T > 600~{\rm GeV}, ~~p_T^j > 30~{\rm GeV}, 
~~ p_T^l > 20~{\rm GeV}; \\ 
R5&:& T_T^m > 200~{\rm GeV}, ~~H_T > 350~{\rm GeV}, ~~p_T^j > 30~{\rm GeV}, 
~~ p_T^l > 20~{\rm GeV}; \\ 
R6&:& T_T^m > 200~{\rm GeV}, ~~H_T > 350~{\rm GeV}, ~~p_T^j > 150~{\rm GeV}, 
~~ p_T^l > 25~{\rm GeV}.
\end{eqnarray}
The discovery domains in $m_{N_1}$ versus $m_\zeta$ are shown in Figs.~8 and 9.

\section{Conclusion}
{\it Question} What does the one Higgs really tell us?  The answer may be that 
flavor and dark matter are also connected and they do so through the one Higgs. 
This new notion extends the scotogenic neutrino mass to all (or most) quark 
and lepton masses, with flavored dark matter.  In this framework, the new 
physics scales responsible for DM and neutrino mass are the same, say 1 TeV. 
If so, its impact is verifiable in the near future at the LHC.

\begin{acknowledgments}
I thank Shinya Kanemura and all the other organizers of HPNP2015 for their 
great hospitality in Toyama.  This work is supported in part by 
U.~S.~Department of Energy under Grant No.~DE-SC0008541.
\end{acknowledgments}

\bigskip 

\end{document}